\documentstyle[12pt,a4wide]{article}
\begin{document}
\input epsf
\newcommand{\be}{\begin{equation}}
\newcommand{\ee}{\end{equation}}
\newcommand{\pr}{\partial}
\newcommand{\ie}{{\it ie }}
\newcommand{\bphi}{\mbox{\boldmath $\phi$}}
\font\mybb=msbm10 at 11pt
\font\mybbb=msbm10 at 17pt
\def\bb#1{\hbox{\mybb#1}}
\def\bbb#1{\hbox{\mybbb#1}}
\def\bZ {\bb{Z}}
\def\bR {\bb{R}}
\def\bE {\bb{E}}
\def\bT {\bb{T}}
\def\bM {\bb{M}}
\def\bC {\bb{C}}
\def\bA {\bb{A}}
\def\bP {\bb{P}}
\def\e  {\epsilon}
\def\bbC {\bbb{C}}
\newcommand{\CP}{\bC \bP}
\newcommand{\CPP}{\bbC \bbb{P}}
\renewcommand{\theequation}{\arabic{section}.\arabic{equation}}
\newcommand{\news}{\setcounter{equation}{0}}
\newcommand{\I}{{\cal I}}
\newcommand{\HH}{\bb{H}}
\newcommand{\bx}{{\bf x}}

\title{\vskip -70pt
\begin{flushright}
{\normalsize \sl To appear in the Milan Journal of Mathematics} \\
\end{flushright}\vskip 50pt
{\bf \large \bf POLYHEDRA IN PHYSICS, CHEMISTRY AND GEOMETRY 
\footnote{Based on the Leonardo da Vinci Lecture given by the first
author in Milan in October 2001.}}\\[30pt]
\author{Michael Atiyah$^{\ \dagger}$ and Paul Sutcliffe$^{\ \ddagger}$
\\[10pt]
\\{\normalsize $\dagger$ {\sl School of Mathematics,}}
\\{\normalsize {\sl University of Edinburgh,}}
\\{\normalsize {\sl King's Buildings, Edinburgh EH9 3JZ, U.K.}}
\\{\normalsize {\sl Email : atiyah@maths.ed.ac.uk}}\\
\\{\normalsize $\ddagger$  {\sl Institute of Mathematics,}}
\\{\normalsize {\sl University of Kent,}}\\
{\normalsize {\sl Canterbury, CT2 7NZ, U.K.}}\\
{\normalsize{\sl Email : P.M.Sutcliffe@kent.ac.uk}}\\}}
\date{February 2003}
\maketitle

\begin{abstract}
\noindent 
In this article we review some
problems in physics, chemistry and mathematics that lead naturally
to a class of polyhedra which include the Platonic solids.
Examples include the study of electrons on a sphere, cages of carbon atoms,
 central configurations of gravitating point particles,
rare gas microclusters,
soliton models of nuclei, magnetic monopole scattering and 
geometrical problems concerning point particles.
\end{abstract}
\newpage

\section{Introduction}\news\label{sec-intro}
\ \quad Polyhedra, particularly the Platonic solids, have been studied by
geometers for thousands of years. Furthermore, finding physical applications
of polyhedra is a similarly ancient pursuit.
 Plato was so captivated by the 
perfect forms of the five regular solids that in his dialogue
{\em Timaeus} he associates them with what, at that time, were believed 
to be the basic elements of the world, namely,
earth, fire, air, water and ether. Kepler also attributed cosmic
significance to the Platonic solids.
In his book {\em Mysterium Cosmographicum} he presents
a version of the solar system
as nested Platonic solids, the radii of the intervening 
concentric spheres being related to the orbits of the planets.
This model had the compelling feature that the existence of only
five Platonic solids explained why there were only the six planets
known at that time. The models of both Plato and Kepler are, 
of course, entirely false, but more promising applications have
since come to light.
 
Recently, there has been an increasing interest in a 
class of polyhedra, which include the Platonic solids,
because they arise naturally in a number of diverse problems in physics,
 chemistry, biology and a variety of other disciplines. 
In this article we shall describe this class of polyhedra and some
problems in physics, chemistry and mathematics which lead naturally
to their appearance. 
As we do not discuss any applications to biology the interested reader
may wish to consult the book by Weyl \cite{Weyl}, 
for a general discussion of symmetry in the natural world.
For a more technical example, a particularly interesting 
application of polyhedra in biology is provided by the structure of
spherical virus shells, such as HIV which is built around a 
trivalent polyhedron with icosahedral symmetry \cite{Ner,For}.

We begin, in the next Section, by recalling some details of the
 Platonic solids, before going on to discuss more general polyhedra.
In the following Sections we then describe some applications in which these
polyhedra arise. First, we deal with systems described by point particles.
The particles
have forces that act between them and we are interested in 
equilibrium states (usually stable ones) in which the particles
sit at the vertices of certain polyhedra. To begin with we deal
with 2-particle forces, such as the inverse square law, which describes
both gravitational attraction and the repulsion of identical electric charges.
To obtain polyhedral solutions with a single 2-particle force requires
 either a constraint on the particles, or an additional force, in 
order to balance the tendency to collapse or expand 
under the action of a single force. We describe examples of both 
situations and then turn to some geometric multi-particle forces which
solve the scaling problem in a novel way, by being scale invariant.
Finally, in Section \ref{sec-skyrmions}, we move on from point particles and 
turn to classical field theories, discussing Skyrmion solutions
and how they relate to polyhedra.

\section{Platonic Solids and Related Polyhedra}
\begin{figure}[ht]
\begin{center}
\leavevmode
\ \hskip 0cm
\ \vskip 0cm
\hbox{
\epsfxsize=15cm\epsffile{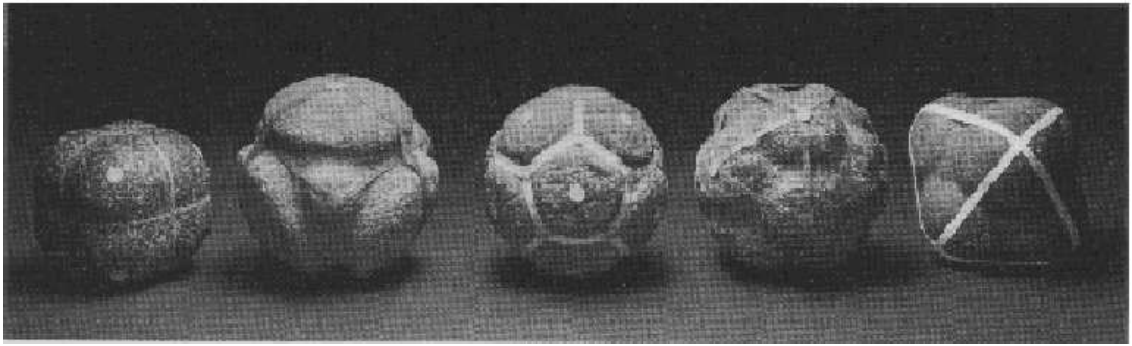}
}
\vskip -0.0cm
\caption{Stone models of the cube, tetrahedron, dodecahedron, icosahedron
and octahedron. They date from about 2000BC  and are kept in the
Ashmolean Museum in Oxford.}
\label{fig-ash}
\end{center}
\end{figure}

\news\label{sec-platonic}
\ \quad
A Platonic, or regular, solid is a polyhedron whose faces are identical
regular polygons with all vertex angles being equal. There are precisely
5 Platonic solids,
the tetrahedron, octahedron, cube, icosahedron and dodecahedron.
In Table \ref{tab-platonic} we list the number of vertices, faces
and edges of the Platonic solids. Although they are termed Platonic solids
there is convincing evidence that
they were known to the Neolithic people of Scotland at least a thousand years
before Plato, as demonstrated by the stone models pictured in 
fig.~\ref{fig-ash} which date from this period and are kept in the
Ashmolean Museum in Oxford.
\begin{table}
\centering
\begin{tabular}{|c|c|c|c|}
\hline
Polyhedron & $V$ & $F$ & $E$\\\hline
tetrahedron & 4 & 4 & 6\\
octahedron & 6 &8 & 12\\
cube & 8 & 6 & 12\\
icosahedron & 12 & 20 & 30\\
dodecahedron & 20 & 12 & 30\\
\hline
\end{tabular}
\caption{The number of vertices $V$, faces $F$ and edges $E$
of the 5 Platonic solids.}
\label{tab-platonic}
\end{table}

The tetrahedron, octahedron and icosahedron all have triangular faces
and are therefore examples of deltahedra. A deltahedron is a polyhedron
with only triangular faces, and is called regular if all the triangles
are equilateral. There are only 8 convex regular deltahedra, and 
the number of their vertices covers the range from 4 to 12, excluding 11.
It is perhaps of interest to note that in 1813 Cauchy proved that
any convex polyhedron is rigid, but it is known that there are
non-convex polyhedra which are not rigid \cite{Con}.
The cube has square faces and the faces of a dodecahedron are regular
pentagons. 
The tetrahedron, cube and dodecahedron are trivalent polyhedra, 
which means that precisely 3 edges meet at every vertex.

For any polyhedron the number of 
vertices $V$, faces $F$ and edges $E,$ must satisfy Euler's formula,
which states that
\be
V+F-E=2.
\label{eulers}
\ee

Given a polyhedron one can construct its dual, which is a polyhedron in 
which the locations of the vertices and face-centers are exchanged.
The octahedron and cube are dual to each other, as are the
icosahedron and dodecahedron. The dual of a tetrahedron is again a 
tetrahedron. From the definitions given above it is clear that the dual
of a deltahedron is a trivalent polyhedron. 

The rotational symmetry groups of the tetrahedron, octahedron and
icosahedron are finite groups of order 12, 24 and 60,
which we denote by $T,O$ and $Y$ respectively. Since the cube is
dual to the octahedron then its symmetry group is also $O,$ and,
again by duality, $Y$ is the symmetry group of the dodecahedron.
Although there are only 5 Platonic solids there are an infinite number of
polyhedra which are symmetric under the Platonic symmetry groups $T,O,Y.$

\section{Particles on a Sphere}
\news\label{sec-sphere}\ \quad
In this Section we consider a set of point particles
whose positions are constrained in some way; the most natural physical 
constraint being that the particles are restricted to the surface
of a unit sphere.

The first problem we discuss is a model for electrons on a sphere.
This is often known as the Thomson problem, even though the
problem which J.J. Thomson explicitly posed 
\cite{JJT} is a little different to this one,
and is described in the next Section. The origin of Thomson's name
being attached to the study of electrons on a sphere appears to
be the paper by Whyte \cite{Wh}, in which the term Thomson problem
is used in this context. 

Consider $n$ particles, with positions ${\bf x}_i$, $i=1,...,n,$
and constrained to lie on the surface of the unit sphere, 
so that $|{\bf x}_i|=1$ for all $i.$ The particles have unit
electric charge and therefore the total Coulombic energy
(in suitable units) is given by
\be
E_1=\sum_{i}^n\sum_{j<i} \frac{1}{|{\bf x}_i-{\bf x}_j|} .
\label{coul}
\ee
The Thomson problem is, for a given $n$, to find the positions
${\bf x}_i$ so that this energy is minimal. The resulting points
are then in equilibrium under the action of their mutual Coulomb
repulsions, the only net forces acting on each particle being normal
to the sphere upon which they are constrained to lie.

Obviously, if there are only two particles then they will sit at
any pair of antipodal points on the sphere. For $n=3$ one finds
that they sit at the vertices of an equilateral triangle on a great
circle. For $n=4,6,12$ the points sit at the vertices of the
Platonic solids, namely the tetrahedron, octahedron and icosahedron, 
respectively. For all $n\ge 4$ the points sit at the vertices of
a polyhedron, which is generically a deltahedron.

\begin{figure}[ht]
\begin{center}
\leavevmode
\ \hskip 0cm
\ \vskip 0cm
\hbox{
\epsfxsize=10cm\epsffile{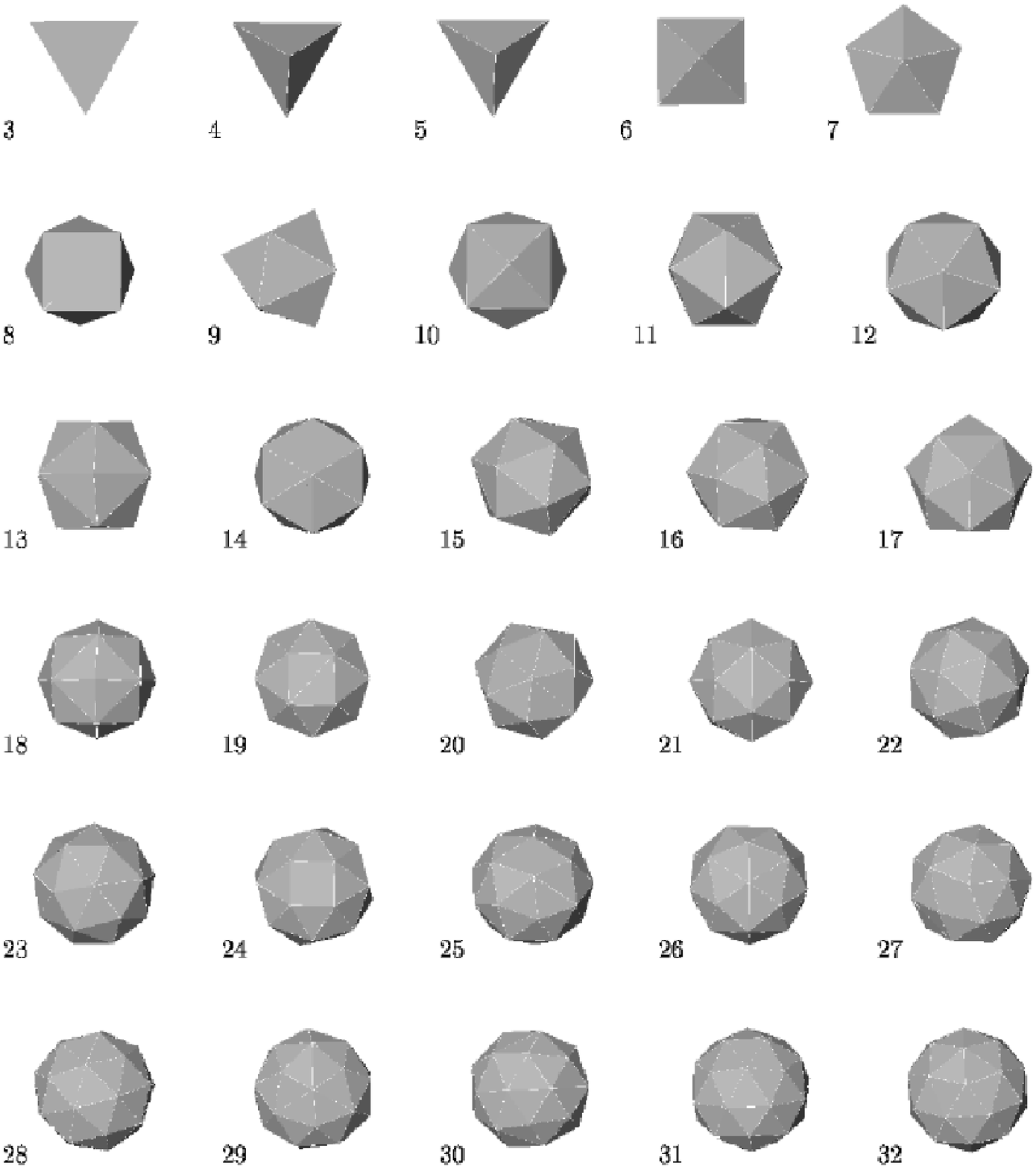}
}
\vskip -0.0cm
\caption{Polyhedra with vertices located at the $n$ points 
on the sphere which
minimize the energy of the Thomson problem for $3\le n\le 32.$ }
\label{fig-electrons}
\end{center}
\end{figure}

In fig.\ref{fig-electrons} we display the associated polyhedra
for $n\le 32.$ Note that 
$n=8$ is the first example in which the polyhedron is not a deltahedron,
it being a square anti-prism, obtained from a cube
by rotating the top face by $45^\circ$ relative to the bottom face. 
This example demonstrates a general feature that the most symmetric 
configurations are not automatically those of lowest energy, since
the more symmetric cube is not favoured for 8 points. Similarly,
the dodecahedron is not obtained for 20 points, since it is
far from possessing the deltahedral property which is clearly favoured.

\begin{table}
\centering
\begin{tabular}{|r|c|}
\hline
$n$ & $G$\\
\hline
   2&           $D_{\infty h}$\\  
   3&           $D_{3h}$\\        
   4&          $T_d$\\           
   5&         $D_{3h}$\\        
   6&          $O_h$\\          
   7&          $D_{5h}$\\       
   8&        $D_{4d}$\\        
   9&           $D_{3h}$\\     
  10&        $D_{4d}$\\    
  11&          $C_{2v}$\\  
  12&         $Y_h$\\      
  13&        $C_{2v}$\\    
  14&         $D_{6d}$\\   
  15&         $D_3$\\      
  16&          $T$\\       
  17&        $D_{5h}$\\    
  18&         $D_{4d}$\\   
  19&          $C_{2v}$\\  
  20&      $D_{3h}$\\      
  21&         $C_{2v} $\\  
  22&       $T_d$\\        
  23&       $D_3$\\        
  24&       $O$\\          
  25&         $C_{1h}$\\   
  26&        $C_2$\\       
  27&          $D_{5h}$\\  
  28&         $T$\\        
  29&        $D_3$\\       
  30&         $D_2$\\      
  31&         $C_{3v}$\\   
  32&         $Y_h$\\      
\hline
\end{tabular}
\caption{The symmetry group $G$ of the configuration of $n$ points which
minimizes the energy of the Thomson problem, for $2\le n\le 32$.}
\label{tab-sym}
\end{table}

In Table \ref{tab-sym} we present the symmetries of the minimal energy 
configurations for $n\le 32,$ obtained from numerical simulations.
To obtain the global minimum for large values of $n$ is a
notoriously difficult numerical problem, due to the fact that 
there is a rapid growth of local minima with increasing $n.$
However, the use of modern computers
has provided numerical results for $n\le 200,$ and their symmetries
have been identified \cite{RSZ,EH}. The numerical methods used on
this problem are numerous and varied, including multi-start conjugate
gradient algorithms, simulated annealing and genetic codes. In fact,
because of its complexity and richness,
the Thomson problem is often used as a test problem in evaluating the 
efficiency of new numerical minimization algorithms.

 In case the reader is not familiar
with the notation used in Table \ref{tab-sym} for the symmetry groups
 we briefly recount the main details here.
The dihedral group $D_k$ is obtained from the cyclic group of
order $k$, $C_k$, by the addition of a $C_2$ axis which is orthogonal
to the main $C_k$ symmetry axis. 
The group $D_k$ can  be extended by the addition of a reflection
symmetry in two ways: by including a reflection in the plane
perpendicular to the main $C_k$ axis, which produces the group $D_{kh}$
or, alternatively, a
reflection symmetry may be imposed in a plane which contains the main
symmetry axis and bisects the $C_2$ axes, which results
in the group $D_{kd}.$ In the same way as for the dihedral groups
the Platonic groups $T,O,Y$, which we introduced earlier, may also be
 enhanced by reflection symmetries,
again denoted by the subscripts $h,d.$ The addition of a subscript
$h$ to a cyclic group denotes a horizontal reflection symmetry,
but a vertical reflection plane is denoted by a subscript $v.$

An obvious extension of the Thomson problem is to replace the Coulomb 
interaction (\ref{coul}) by a more general power law
\be
E_p=\sum_{i}^n\sum_{j<i} \frac{1}{|{\bf x}_i-{\bf x}_j|^p}.
\label{gen}
\ee
Again polyhedral solutions are obtained, but the symmetry and
 structure of the minimal energy configuration generally has an
interesting  non-trivial behaviour as a function of $p$ (and $n$) \cite{MKS}.

In the limit as $p\rightarrow\infty$ the generalized problem (\ref{gen})
is equivalent to the Tammes problem, named after the Dutch botanist
who posed the problem in connection with the study of pores
in spherical pollen grains \cite{Ta}.
The Tammes problem is to determine
the configuration of $n$ points on a sphere so that the minimum distance
between the points is maximized. 
This is equivalent to asking how to pack $n$ identical circular caps on the
 surface of a sphere so that the diameter of the caps is as large as possible.
As expected, polyhedral solutions are 
also obtained for the Tammes problem, but, perhaps surprisingly,
 for $n>6$ the numerical results
suggest that the only configuration which is a common solution
of the Coulomb and Tammes problem is the icosahedral arrangement for $n=12$
\cite{EH}. 

The final example that we mention for points on a sphere is a
problem in discrete geometry which was first posed about fifty years
ago by Fejes T\'oth \cite{FT}. This problem is to maximize the sum of
 the mutual separations, or equivalently to minimize the energy
\be
E_{-1}=-\sum_i^n \sum_{j<i}|{\bf x_i}-{\bf x_j}|.
\label{Toth}
\ee
The solutions of this problem are also polyhedral and usually,
 though not always, have the same symmetry group and combinatorial type as 
the solutions of the Thomson problem \cite{RSZ}. For example, the
 symmetry groups
listed in Table \ref{tab-sym}, for $2\le n\le 32,$ are also the symmetry
groups of this problem,  excepts for the cases $n=7$  and $n=29.$
For the Thomson problem the $n=7$ polyhedron is a regular pentagonal
bipyramid (see fig.~\ref{fig-electrons}) 
but for this problem the polyhedron is a symmetry breaking
perturbation of this bipyramid, having only a $C_2$ symmetry with
2 antipodal points and the remaining 5 points sprinkled around
an equatorial band. The symmetry of the extremal configuration
of 29 points is $C_{2v}$ rather than the $D_3$ symmetry of the Thomson 
problem.

There are a number of theorems proved about the extrema of the
 energy (\ref{Toth}) and in particular there is the lower bound \cite{Al}
for the minimum 
\be
E > \frac{1}{2}-\frac{2}{3}n^2,
\label{Tothbound}
\ee
which is also a reasonable estimate of the true minimal energy.

\section{Central Configurations and Related Problems}
\news\label{sec-central}\ \quad
In this Section we consider point particles whose positions are
unconstrained, but which have an additional central force
 in order to balance a 2-particle interaction.

First, we consider central configurations, which are minima of the
energy 
\be
E= \sum_i^n \sum_{j <  i} {1  \over |{\bf x}  _i -{\bf x}_j |}
+\sum_i^n { 1\over 2}  {\bf x}^2_i,
\label{central}
\ee
where each point ${\bf x}_i$, for $i=1,...,n$, is now unconstrained
and so may take any value in $\bR^3.$ The variation of this energy
yields the equilibrium equations
\be
{\bf x}_i + \sum_{j\ne i} {
{\bf x}_j -{\bf x}_i \over |{\bf x}  _i -{\bf x}_j |^3 }=0,
\label{centraleom}
\ee
 in which there is a balance between
a repulsive inverse square force between the particles and a linear
attraction to the origin.  

One physical interpretation of the energy (\ref{central}) 
arises when the inverse square 
force is thought of as an electrostatic repulsion between like charges 
and the linear force
as an attraction due to a uniform background of the opposite
charge. In this guise the problem originally arose in J.J. Thomson's
static Plum Pudding model of the atom~\cite{JJT} in which the positive electric
charge is smeared out into a uniform ball (the pudding) while the
negatively charged electrons correspond to the plums. Although
Rutherford's experiments conclusively  demonstrated that this model is
not relevant as a theory of atomic structure, it nevertheless
continues to offer insights into the structure of metals
(with the role of positive and negative charges interchanged)
 and other condensed matter systems and is often referred to as 
the One Component Plasma (OCP) model~\cite{Bau}, 
or sometimes as classical Jellium.  

An alternative interpretation of the equilibrium equations (\ref{centraleom})
is that it describes a force balance between gravitating point particles,
with an attractive inverse square force law, and a repulsive radial force
proportional to the distance from the origin. 
The repulsive linear force
is such as arises in theories with a cosmological constant
and also arises naturally if a time dependent
homothetic ansatz is used in Newton's equations of motion. 
For the gravitational interpretation the appropriate potential function is
$-E$, so minimizing $E$ does provide critical configurations, that is, they
are solutions of the equilibrium equations (\ref{centraleom}),
but they are not minima of the gravitational energy $-E.$ 
However, the issue of dynamical stability is more
complicated in the gravitational setting. 
The actual particles are not static, because of the time dependent
homothetic ansatz, so the instability of these solutions requires
more than just the existence of a negative mode for the energy $-E,$
and is by no means obvious.

It turns out that the electrostatic interpretation provides a useful
insight into the properties of the configurations which minimize
the energy (\ref{central}). For example, a lower bound on the energy 
can be obtained
by considering the packing of $n$ non-overlapping Thomson type Hydrogen
atoms, that is, spheres with a single point charge at their centre but with 
zero total charge inside the sphere. This bound is given by \cite{LN,BGS}
\be
E\ge \frac{9}{10}n(n^{2/3}-1)
\ee
and is a reasonably good estimate of the true minimal energy, 
indicating that the sphere packing assumption under which it is derived
is a good description of the true minimizing configuration.
\begin{figure}[ht]
\begin{center}
\leavevmode
\ \hskip 0cm
\ \vskip 0cm
\hbox{
\epsfxsize=10cm\epsffile{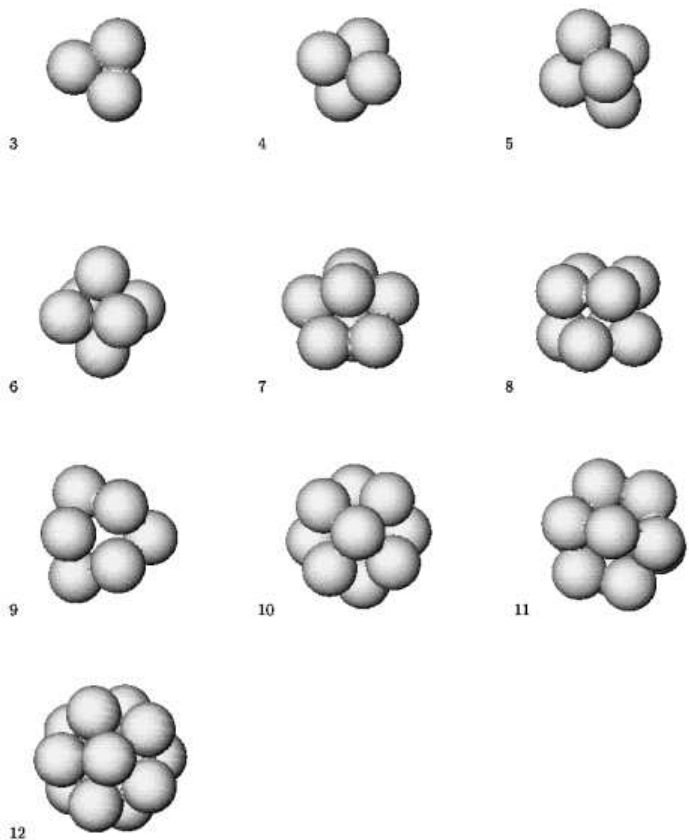}
}
\vskip -0.0cm
\caption{For $3\le n\le 12$ the minimal energy central configurations
of $n$ points are displayed
by plotting spheres of diameter $d=1.65$ around each of the points. }
\label{fig-central}
\end{center}
\end{figure}
For $3\le n\le 12$ the minimal energy configurations are displayed
in fig.~\ref{fig-central} by plotting spheres of diameter $d=1.65$ around
each of the points. 
This illustrates the sphere packing nature of the solutions.
The numerical value of 1.65 for the diameter $d$ was determined using
statistical data from configurations with large values of $n,$ though
analytic considerations lead to a similar value \cite{BGS}. For $n\le 12$
all the points lie on, or very close to, the surface of a sphere and
the symmetry and combinatoric type of each of the configurations matches
that for the Thomson problem.
 Thus for $n<13$ the polyhedra associated
with this problem are essentially those of the Thomson problem.
However, for $n\ge 13$ this is no longer true and not all the points
lie close to the surface of a single sphere. For example, at $n=13$ there
are 12 points on the vertices of an icosahedron and an additional point at
the origin. This may be thought of as a 2-shell structure, with the first
shell containing a single point and the second shell containing 12 points.
A 2-shell structure persists for all $13\le n\le 57$ and $n=60,$ with
the arrangement of points within each shell resembling that for the
Thomson problem. For example, for $n=32$ the inner shell contains 
4 points on the vertices of a tetrahedron and the 28 points in the 
outer shell are also arranged in a tetrahedrally symmetric configuration,
as suggested by the symmetry of the Thomson problem in Table \ref{tab-sym}.
These two tetrahedral arrangements are appropriately aligned so that the
whole structure has tetrahedral symmetry, and this gives the $n=32$
configuration a particularly low energy. 

\begin{figure}[ht]
\begin{center}
\leavevmode
\ \hskip 0cm
\ \vskip 0cm
\hbox{
\epsfxsize=15cm\epsffile{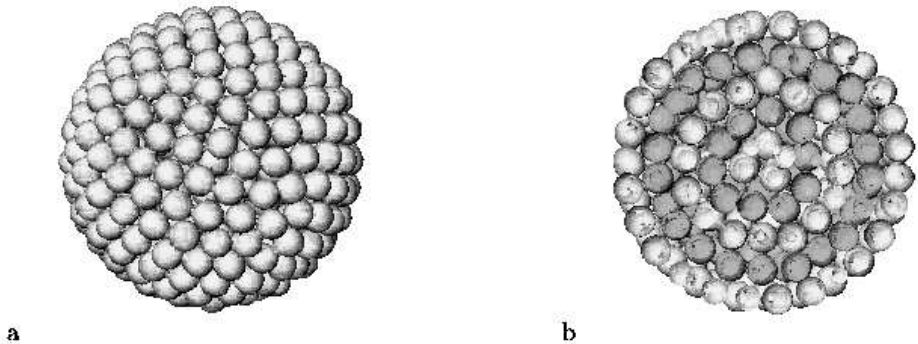}
}
\vskip -0.0cm
\caption{(a) A sphere packing presentation of a central configuration
with 500 points. (b) A slice through the centre of this
configuration, highlighting the 5 shells by alternately shading the shells
light and dark grey.}
\label{fig-500}
\end{center}
\end{figure}
The number of shells increases with $n$ and so for $n$ up to around 1000
points we may picture the configuration as points which sit at the vertices 
of a set of nested polyhedra. In fig.~\ref{fig-500}a we display,
using the same sphere packing presentation as in fig.~\ref{fig-central},
 a low energy local minimum configuration containing 500 points,
which is a good candidate for the global minimum.
In fig.~\ref{fig-500}b we present a slice through the centre of this
configuration, highlighting the 5 shells by alternately shading the shells
light and dark grey. At around 1000 points the
shell-like structure begins to break down, with the distinction between
the shells becoming blurred. For larger values of $n$ the configuration
resembles a random close packing of hard spheres, similar to Bernal's
model of liquids. The configuration tends towards having a constant density,
with 2-point and 3-point probability distributions agreeing to a high precision
with those of a uniform distribution of points inside a ball \cite{BGS}.
There is numerical evidence \cite{Tot} for a transition,
at some value between 11000 and 15000 points, to a body centred cubic structure
as the minimal energy arrangement.

As we have seen, the points which minimize the energy (\ref{central})
lie on the vertices of a single polyhedron only for $n\le 12,$ and for larger
$n$ nested polyhedra appear since the points do not all lie close to
the surface of a single sphere. However, there is a variation of the
central configuration energy which, for all values of $n$,
 does appear to have all the points lying
close to the surface of a sphere \cite{BGRS}.
This involves replacing the Coulomb energy, the first term in (\ref{central}),
by the energy (\ref{Toth}), discussed in Section \ref{sec-sphere}
when studying points on the sphere which maximize the sum of their mutual
 separations. The modified energy is given by
\be
E= -\sum_i^n \sum_{j <  i}|{\bf x}  _i -{\bf x}_j |
+\sum_i^n { 1\over 2}  {\bf x}^2_i,
\label{monopole}
\ee
where each of the $n$ points ${\bf x}_i$ is again unconstrained.

For the minimal energy configurations all $n$ points lie on, or very
close to, the surface of a sphere whose radius grows linearly with $n$ and 
is well approximated by the formula $r=\frac{2n}{3}-\frac{1}{2n}.$
The arrangement of the points is essentially the same as for the
energy (\ref{Toth}), when the points are contrained to lie precisely on
this sphere. This observation allows the use of the estimate (\ref{Tothbound})
to provide the approximation \cite{BGRS}
\be
E=-\frac{2}{9}n^3 +\frac{1}{3}n-\frac{1}{8}
\ee
which is a good estimate of the true minimal energy value.

The energy function (\ref{monopole}) appeared in the study of monopole
scattering \cite{BGRS}, as we now briefly describe. BPS monopoles are
topological solitons in three space dimensions and the equation describing
a static $n$-monopole solution has a moduli space of dimension $4n.$
The $4n$ parameters may be thought of as describing a position and phase
for each of the $n$ monopoles and the low energy scattering of monopoles can
be approximated by geodesic motion on the moduli space \cite{Ma}.
For $n$ fundamental monopoles in an $SU(n+1)$ gauge theory 
the metric on the moduli space is the  
hyperk\"ahler Lee-Weinberg-Yi metric \cite{LWY},
which is known explicitly. There are geodesics which describe a time
dependent scaling solution, providing the monopole positions satisfy
a set of algebraic constraints. These constraints are precisely the
equations satisfied by the critical points of the energy function
(\ref{monopole}), and so minima of this energy yield geodesics,
which have an interpretation in terms of monopole scattering, with
the points identified as the positions of the monopoles.
These solutions describe the scattering of $n$
monopoles which  lie on the vertices of a bouncing polyhedron;
the polyhedron contracts from infinity to a point, representing the spherically
symmetric $n$-monopole, and then expands back out to infinity. 

\section{Lennard-Jones Clusters}
\news\label{sec-LJ}\ \quad
In the previous Section we studied unconstrained 
particles with a repulsive 2-body force and an additional linear force
producing attraction to the origin. In this Section we discuss the
situation in which the particles are acted upon by two 2-body
forces with opposite characters, that is, one force is repulsive
and the other is attractive, leading to a finite non-zero separation
at which two particles feel no mutual force. 

A particularly interesting example is the Lennard-Jones interaction energy   
\be
E=\sum_i^n \sum_{j <  i}\left\{
\frac{1}{|{\bf x}  _i -{\bf x}_j |^{12}}
-\frac{2}{|{\bf x}  _i -{\bf x}_j |^{6}}
\right\}
\label{LJ}
\ee
which is a good model for rare gas microclusters \cite{Ho}.
A microcluster is an aggregate of a small number of atoms, where
an appreciable number of the atoms reside at the surface of the structure.
For microclusters of rare gases the most relevant force which binds
the atoms is a weak van der Waals force, and this is well approximated by
the classical Lennard-Jones energy.

In (\ref{LJ}) we have scaled the length and energy units so that two
atoms a unit distance apart feel no mutual force. For $3\le n \le 7$
the symmetry of the energy minimizing configuration agrees with that
shown in Table \ref{tab-sym}, with the atoms sitting at the vertices
of an equilateral triangle, tetrahedron, triangular bipyramid, octahedron
and pentagonal bipyramid respectively, in qualitative agreement with
 the polyhedra displayed in fig.~\ref{fig-electrons}. However, for $n\ge 8$
the structures which emerge are, generically, quite different from 
the problems we have considered so far, although they are connected
to polyhedra in a slightly different way.

\begin{figure}[ht]
\begin{center}
\leavevmode
\ \hskip 0cm
\ \vskip -2cm
\hbox{
\epsfxsize=6cm\epsffile{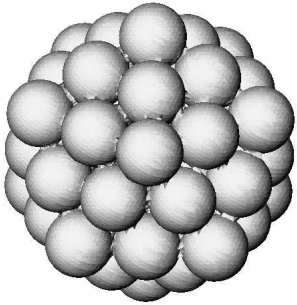}
}
\vskip -0.0cm
\caption{The Lennard-Jones cluster for 55 atoms. Each atom is
represented by a sphere of unit diameter.}
\label{fig-LJ55}
\end{center}
\end{figure}

The minimal energy arrangement of atoms is, almost always, obtained
from a filling and partial filling of the vertices of Mackay icosahedra.
Consider a sphere surrounded by 12 identical spheres on the vertices
of an icosahedron. As pointed out by Mackay \cite{Mackay} one may
consider the icosahedron as the first shell in a sequence of shells, 
$s=1,2,3,...$. Each shell contains $10s^2+2$ spheres and is packed around
the previous one in a way that preserves the icosahedral symmetry. 
Clearly this leads to a sequence of numbers 13,55,147,... in which
an outer shell and all inner shells are completely filled. These
complete Mackay icosahedra are the minimal energy Lennard-Jones 
configurations for these special values of $n.$
In fig.~\ref{fig-LJ55} we present the $n=55$ Lennard-Jones cluster
by plotting spheres of unit diameter around each of the 55 atoms.
The outer shell containing 42 atoms is visible and surrounds 12
atoms on the vertices of an icosahedron with a single atom at its
centre. Remarkably, as first demonstrated by Northby \cite{No},
partially filling the outer shell and completely filling the inner
shells often produces the minimal energy configuration. For example,
the minimal energy arrangement for 12 atoms is obtained from the $n=13$
Mackay icosahedron by removing an atom from a vertex of the icosahedron,
not by removing the atom at the centre. Thus the symmetry of the
$n=12$ minimal energy configuration is only $C_{5v}$ and not the
more symmetric icosahedral arrangement obtained by placing all 12 atoms on the
vertices of an icosahedron. 

There are a few values of $n$ for which the minimal energy configuration
is not based on Mackay icosahedra \cite{Wales} and these are difficult to find
numerically because of the enormous number of local minima which exist
for this problem. For example, even at $n=13$ numerical methods have
computed 1467 different local minima \cite{DMW}.

Using electron diffraction techniques  
experimental studies \cite{Fa} of argon microclusters in nozzle beams have 
revealed a remarkable agreement with the icosahedral based structures
predicted by minimization of the simple Lennard-Jones energy. 
Also, intensity anomalies coincide with the particularly stable 
atom numbers associated with complete Mackay icosahedra.

\section{Geometric Energies}
\news\label{sec-geometric}\ \quad
So far we have seen two situations in which minimal energy
configurations have the particles located at the vertices of a single 
polyhedron. The first is where the particles are constrained to lie
on the surface of a sphere and the second is the energy (\ref{monopole})
in which a 2-body repulsive force for unconstrained particles 
is balanced by an attraction towards the origin. 
In this Section we describe some geometric energies with
multi-particle forces which produce polyhedral minimizers and
solve the scaling problem in a novel way, by being scale invariant.

To construct our first geometric energy consider $n$ distinct 
ordered points, ${\bf x}_i\in\bR^3$ for $i=1,...,n.$
 For each pair $i\ne j$ define the unit vector
\be
{\bf v}_{ij}=\frac{\bx_j-\bx_i}{|\bx_j-\bx_i|} 
\label{unitv}
\ee
giving the direction of the line joining $\bx_i$ to $\bx_j.$
Now let $t_{ij}\in\CP^1$ be the point on the Riemann sphere associated with
the unit vector ${\bf v}_{ij}$, via the identification 
$\CP^1\cong S^2,$ realized as stereographic projection. 
Next, set $p_i$ to be the polynomial in $t$ with
roots $t_{ij}$ ($j\ne i$), that is
\be
p_i=\alpha_i\prod_{j\ne i}(t-t_{ij})
\label{defp}
\ee
where $\alpha_i$ is a certain normalization coefficient (see ref.\cite{AS}
for full details).
In this way we have constructed $n$ polynomials which all have degree
$n-1,$ and so we may write 
$p_i=\sum_{j=1}^n d_{ij}t^{j-1}.$ 
Finally, let $d$ be the 
$n\times n$ matrix with entries $d_{ij},$ and let $D$ be its
determinant
\be
D({\bf x}_1,...,{\bf x}_n)=\mbox{det}\ d.
\ee
This geometrical construction is relevant to the Berry-Robbins
problem \cite{BR}, which is concerned with specifying how a spin
basis varies as $n$ point particles move in space, and supplies
a solution provided it can be shown that $D$ is always non-zero.
For $n=2,3,4$ it can be proved that $D\ne 0$ \cite{At,EN} and
 numerical computations \cite{AS} suggest that $|D|\ge 1$ for all $n,$
with the minimal value $|D|=1$ being attained by $n$ collinear points.
  
The geometric energy is the $n$-point energy defined by
\be
E=-\log |D|,
\label{geom}
\ee
so minimal energy configurations maximize the modulus of the determinant.

This energy is geometrical in the sense that it only depends on the
directions of the lines joining the points, so it is translation, rotation 
and scale invariant. Remarkably, the minimal energy configurations,
studied numerically \cite{AS} for all $n\le 32,$ 
 are essentially the same as those for the Thomson problem displayed in
fig.~\ref{fig-electrons}. The points lie on, or very close to, the surface
of a sphere, though of course the radius and centre of the sphere are
now arbitrary. The symmetries and combinatorial types agree in all cases and
plots of the associated polyhedra cannot be distinguished from those
 in fig.~\ref{fig-electrons}. It is expected that the
points lie close to the surface of a sphere for all values of $n$, though
the numerical evidence is currently only available for $n\le 32.$
It would be interesting to see if the similarity with the Thomson 
problem continues for larger values of $n,$ or whether some
qualitative differences arise for $n$ sufficiently large.

A fairly accurate approximation to the minimal energy value is obtained
using the quadratic fit
\be
E(n)=-an^2+bn+4a-2b
\ee
where $a=0.143$ and $b=0.792.$ 

 The geometrical energy (\ref{geom}) may
be interpreted as a volume in a higher dimensional space \cite{AS}
and this point of view may provide a new perspective on configurations
in the Thomson problem.

Often a complicated multi-particle energy function can be expanded
as a sum of pure $k$-particle energies, starting with $k=2,$ which may
be the dominant contribution. For just 2 particles the geometrical energy
 (\ref{geom}) is independent of their positions, so an expansion in terms
of $k$-particle energies must start with $k=3.$
For 3 points the energy (\ref{geom}) may be written as \cite{At}
\be
E^{(3)}=
-\log\{\frac{3}{4}+\frac{1}{4}(\cos\theta_1+\cos\theta_2+\cos\theta_3)\}
\label{e3}
\ee
where $\theta_1,\theta_2,\theta_3$ are the angles in the triangle formed
by the 3 points. In this form it is clear that the minimal value of
$E^{(3)}$ is $-\log(9/8),$ obtained by the equilateral triangle 
$\theta_1=\theta_2=\theta_3=\pi/3.$ The fact that this
3-particle energy might be the dominant part of the full energy (\ref{geom})
motivates the study of the second geometrical energy
\be
E_\Delta=\frac{-6}{n(n-1)(n-2)}\sum_{i>j>k}\log
\{\frac{3}{4}+\frac{1}{4}(\cos\theta_1+\cos\theta_2+\cos\theta_3)\}
\label{edelta}
\ee
where $\theta_1,\theta_2,\theta_3$ are the angles in the triangle formed
by the 3 points $\bx_i,\bx_j,\bx_k.$ The sum runs over all triples of points
and the prefactor counts the number of triples, so that the energy is the
average 3-particle energy per triangle. A numerical study \cite{AS}
of the minimal
energy configurations for the triangular energy (\ref{edelta}) confirms
that it reproduces the arrangements of points for the full energy
(\ref{geom}) to a good accuracy, with the positions of all the particles
differing by less than $1\%$ in all cases.

It is intriguing that the 3-particle energy (\ref{e3}) has arisen
in a different physical context \cite{Cal}. This concerns quantum
many-body problems for which eigenstates can be found explicitly. It turns
out that the addition of precisely this 3-body interaction allows the
explicit computation of some eigenstates (including the ground state)
of less tractable Hamiltonians with only 2-body forces.

Minimal energy configurations for the multi-particle energy (\ref{geom}) 
may also be studied for $n$ points in the plane. 
The results for $2\le n\le 15$ are perhaps not surprising, with the
$n$ points lying on a circle and forming the vertices of a regular 
$n$-gon. However, for $n=16$ the minimal energy configuration consists
of a regular 15-gon on a circle and a single point at the origin.
The pattern of an $(n-1)$-gon plus one point at the centre continues
until $n=23,$ at which point the configuration comprizes a 21-gon
plus 2 points placed in the interior of the circle
 equidistant from its centre
 and lying on a diameter. 
At $n=28$ a further point enters the interior of the circle, 
producing a 25-gon with an equilateral triangle inside.
At $n=33$ another point enters the interior of the circle, leading to a 29-gon 
with a square arrangement at the interior.
 Note that the points in the interior
are always arranged in the minimal energy configuration of that
number. 
We refer to the sequence of numbers, $n=16,23,28,33,\ldots$\
at which an additional point enters the interior of the circle,
as the jumping values. 

These planar configurations are qualitatively similar
to the global minima for the Coulomb energy (\ref{coul}) 
for point charges in a disk \cite{Nur}.
The same patterns of points emerge in the Coulomb case,
though the jumping values
are shifted to the sequence $n=12,17,19,22,\ldots .$
Thus the jumps occur earlier and more frequently for the Coulomb energy.
In the Coulomb problem the number of shells increases for even larger
values of $n$ and we expect a similar phenomenon for the planar
geometric energy.

\section{Fullerenes}
\news\label{sec-fullerenes}\ \quad
A particularly interesting class of trivalent polyhedra occur 
with $F$ faces where $F\ge 12.$ They are composed of 12 pentagons
and $F-12$ hexagons. The trivalent property together with Euler's
formula (\ref{eulers}) 
then determines that the numbers of vertices and edges are
given by $V=2F-4$ and $E=3F-6.$ We refer to a polyhedron of this type
as a Fullerene polyhedron, the name being taken from a particular
application which we shall discuss in this Section.
The first Fullerene polyhedron has $F=12$ and is the dodecahedron.
Another interesting Fullerene with icosahedral symmetry occurs
when $F=32$, and is the truncated icosahedron, which can be
obtained from an icosahedron by \lq chopping off\rq\ each 
of the 12 vertices, leaving 12 pentagons and 20 hexagons.
The number of combinatorially different Fullerenes
grows rapidly with $F.$ For example, although the dodecahedron
is the unique Fullerene with $F=12,$ there are already 40 different
Fullerenes with $F=22$ \cite{atlas}.
Clearly, the dual of a Fullerene is a deltahedron in which
either 5 or 6 triangles surround a vertex, these two cases
being termed pentamers and hexamers respectively. 
The polyhedra which have arisen in previous Sections, with vertices
given by the positions of energy minimizing point particles, are
therefore generically dual to Fullerene polyhedra, once the number of particles
is at least 20.

\begin{figure}[ht]
\begin{center}
\leavevmode
\ \hskip 0cm
\ \vskip -0cm
\hbox{
\epsfxsize=6cm\epsffile{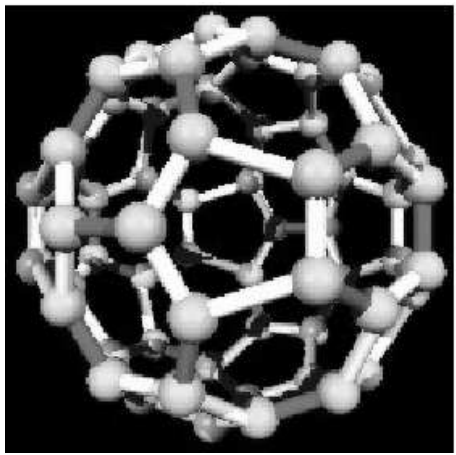}
}
\vskip -0.0cm
\caption{A model of the $C_{60}$ buckyball.}
\label{fig-buckyball}
\end{center}
\end{figure}

Fullerene polyhedra have become of great interest in recent years due
to their unexpected appearance in carbon chemistry. 
Natural carbon can exist in several forms, such as graphite and diamond,
 but there is another form, called Fullerenes.
Fullerenes are large carbon-cage molecules, by far the most common being
the $C_{60}$ buckyball \cite{kroto}, whose discovery was rewarded with the 
1996 Nobel Prize in Chemistry. The carbon atoms sit at the vertices
of a Fullerene polyhedron (as defined above), with the bonds represented by
the edges of the polyhedron. Although carbon is quadrivalent, the polyhedron
is trivalent since each carbon atom is linked to two others by a single
bond and to one further carbon atom by a double bond; the difference in
lengths between a double and single bond is small and so can be neglected.
As an example, 
the 60 atoms in the buckyball
are located on the vertices of a truncated icosahedron, a model of which
is displayed in fig.~\ref{fig-buckyball}.
Other Fullerenes are also common, particularly  $C_{70},C_{76}$ and $C_{84},$
and have been found to exist in
interstellar dust as well as in geological formations on Earth. 
As mentioned above, there is a rapid growth in the number of different
Fullerenes as the number of carbon atoms increases, and it is a challenging
task to catalogue and understand the properties of all possible Fullerenes,
as well as predict, by energy minimization, those most likely to be
 found in experiments. However, simple geometrical rules, such as 
separating the pentagonal faces, appear to play an important role \cite{Zh}.

\section{Skyrmions}
\news\label{sec-skyrmions}\ \quad
In the previous Sections we have dealt with energy minimizing configurations
of point particles, but in this Section we turn to a classical field theory
whose solutions are also related to polyhedra. 

The Skyrme model \cite{Sk} is a nonlinear field theory whose classical
soliton solutions, called Skyrmions,
 are candidates for an effective description of nuclei
with an identification between the number of nucleons and solitons.

The basic field of the model is a 4-component unit vector, 
$\bphi({\bf x})=(\phi_1,\phi_2,\phi_3,\phi_4)$ with $\bphi\cdot\bphi=1.$
The field is defined throughout $\bR^3$ and satisfies the 
boundary condition that $\bphi\rightarrow (0,0,0,1)$ as 
$|{\bf x}|\rightarrow\infty.$ This boundary condition compactifies
space so that topologically $\bphi$ is a map between two 3-spheres.
This map has an integer-valued winding number, $B,$ which is just
the degree of the map, and counts the number of solitons in a given
field configuration. In the application to nuclear physics, $B$ is
identified with baryon number, that is, the number of nucleons.

The image under $\bphi$ of an infinitesimal sphere
of radius $\epsilon$ and centre ${\bf x}\in\bR^3$,
to leading order in $\epsilon$, is an ellipsoid with principal axes
$\epsilon\lambda_1,\epsilon\lambda_2,\epsilon\lambda_3.$

In terms of this deformation the baryon number
can be computed as the integral 
\be
B=\frac{1}{2\pi^2}\int \lambda_1\lambda_2\lambda_3\ d^3x
\label{baryon}
\ee
and the static energy of the model is defined to be \cite{Ma2}
\be
E=\frac{1}{12\pi^2}\int \lambda_1^2+\lambda_2^2+\lambda_3^2
+\lambda_1^2\lambda_2^2+\lambda_2^2\lambda_3^2+\lambda_3^2\lambda_1^2\
d^3 x.
\label{skyrme}
\ee
\begin{figure}[ht]
\begin{center}
\leavevmode
\ \hskip 0cm
\hbox{
\epsfxsize=9cm\epsffile{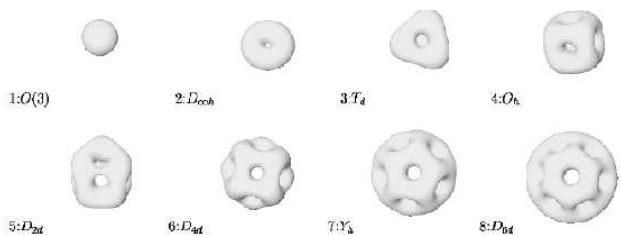}
}
\caption{Baryon density isosurfaces
 for the minimal energy Skyrmions with $1\le B \le 8.$ }
\label{fig-sky1}
\end{center}
\end{figure}
A simple inequality shows that the energy is bounded from below by
the baryon number, that is, $E\ge B,$ which is known as the
 Faddeev-Bogomolny bound. It is easy to see that this bound can
 not be attained for any non-trivial field configuration.
The mathematical problem is to compute, for a given integer $B,$
the minimal energy field configuration with baryon number $B.$
Using numerical methods, and a powerful parallel machine, minimal energy 
Skyrmions have been computed \cite{BS} for all $B\le 22.$
In fig.~\ref{fig-sky1} we present surfaces of constant baryon density
(the integrand in equation (\ref{baryon})) for the minimal energy
Skyrmions with $1\le B \le 8.$ Each Skyrmion is also labelled by
the point symmetry group of the baryon density
 (or equivalently energy density).

\begin{figure}[ht]
\begin{center}
\leavevmode
\ \hskip 0cm
\ \vskip -0cm
\hbox{\epsfxsize=7.5cm\epsffile{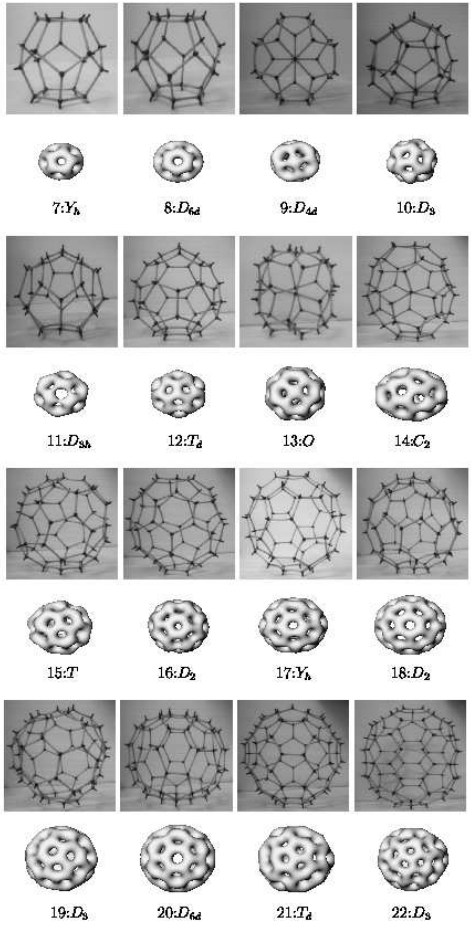}}
\vskip -0.0cm
\caption{Baryon density
isosurfaces for  minimal energy Skyrmions with 
$7\le B \le 22,$ together with models of the associated polyhedra.}
\label{fig-sky2}
\end{center}
\end{figure}

The $B=1$ Skyrmion is spherically symmetric \cite{Sk} but for $B=2$
the Skyrmion has only an axial symmetry \cite{KS,Ve}, with a toroidal
structure for the baryon density isosurface. For $B>2$ the baryon density
is localized around the edges (and especially vertices) 
of a polyhedron, which for $B=3$ and $B=4$
is a tetrahedron and cube respectively \cite{BTC}. 
From fig.~\ref{fig-sky1} it can be seen that each of the polyhedra is trivalent
and contains $2B-2$ faces. This suggests a possible connection with
Fullerene polyhedra which have  $F=2B-2$ faces, or equivalently $V=4B-8$
vertices, for $B\ge 7.$ In fig.~\ref{fig-sky2} we present baryon density
isosurfaces for  minimal energy Skyrmions with 
$7\le B \le 22,$ together with models of the associated
polyhedra. Generically, the polyhedra are of the Fullerene type, for
example,
$B=7$ is the dodecahedron and
 $B=17$ has $V=60$ vertices forming the truncated
icosahedron of the buckyball. Two of the configurations are not of the
Fullerene type ($B=9,13$), as they contain quadrivalent vertices, but they
can be understood as deformations of Fullerene polyhedra \cite{BS}. 
The baryon density isosurfaces displayed in fig.~\ref{fig-sky1} are
qualitatively similar to energy density isosurfaces for BPS magnetic
monopoles \cite{Su}, which are another kind of soliton. The equation
for static monopoles is integrable and this allows exact results to be
proved for monopoles, which can then be used as a guide for studying 
Skyrmions. In particular, there is an exact correspondence between
the moduli space of static $B$-monopoles, and degree $B$ rational maps 
between Riemann spheres \cite{Do,Ja}. This motivated the development of
an approximate ansatz for Skyrmions in terms of rational maps \cite{HMS},
which has proved remarkably successful in capturing both the qualitative and 
quantitative features of Skyrmions. Furthermore, this ansatz depends on
only a (small) finite number of parameters, and induces an energy function
on the moduli space of rational maps. Thus numerical computations
of minimal energy maps can be performed with much less effort than minimizing
the full Skyrme energy function, and yet produce results 
which are in good agreement with the full field theory \cite{BS}.  

\begin{figure}[ht]
\leavevmode
\ \hskip 4cm
\ \vskip -0cm
\hbox{
\epsfxsize=6cm\epsffile{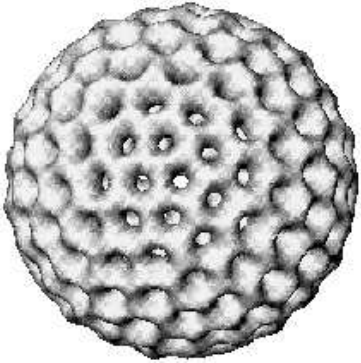}
}
\vskip -0.0cm
\caption{Baryon density isosurface for a $B=97$ Skyrmion with
icosahedral symmetry.}
\label{fig-icos}
\end{figure}

A Skyrmion polyhedron has $2B-2$ faces, so taking its dual yields,
generically, a deltahedron with $n=2B-2$ vertices.
An inspection of the symmetries of the minimal energy Skyrmions, 
presented in the labels in figs.~\ref{fig-sky1} and ~\ref{fig-sky2},
shows that quite often the symmetry of the minimal energy $B$-Skyrmion
agrees with the symmetry of the solution of the Thomson problem
(see Table \ref{tab-sym}) for $n=2B-2$ particles on the sphere.
More precisely, the symmetries match for the two problems in 17 out of
 the 22
cases. Moreover, a closer inspection reveals that in these 17 cases
not only do the symmetry groups match, but the combinatorial types of
the Skyrmion polyhedron and the dual of the Thomson polyhedron are
identical. The 5 examples that do not coincide, $B=5,9,10,19,22$
shows that the topography of the two energy functions is slightly different
 and suggests
that the same factors which determine the polyhedron (or its dual) 
are important, but perhaps with slightly different
weightings. 
The fact that there is often an agreement for the two problems
motivated the study \cite{BHS} of icosahedral Skyrmions with values
of $B$ for which the solution of the Thomson problem with $n=2B-2$ points 
has icosahedral symmetry. The results appear to support a connection
between the two problems and suggests
icosahedral candidates for 
minimal energy Skyrmions with certain large values of $B.$
As an example, the baryon density isosurface shown in fig.~\ref{fig-icos}  
is an icosahedrally symmetric Skyrmion with $B=97.$ It has very
low energy and is a good candidate for the minimal energy 97-Skyrmion.

\section{Conclusion}
\news\label{sec-conclusion}\ \quad
In this review we have discussed a range of problems in physics,
chemistry and mathematics which all yield polyhedral solutions.
Mainly, the problems we have addressed concern minimizing the energy
of point particles, with the result that the particles are arranged
on the vertices of a polyhedron. Sometimes the system involves a
single repulsive 2-particle interaction, and the points are constrained
to lie on the surface of a sphere, so the emergence of a polyhedron is
 perhaps not so surprising -- though the patterns and symmetries of
the polyhedra often are. Perhaps more surprising, is the emergence
of polyhedra when no constraints are placed on the particle positions,
and yet the particles continue to be arranged on, or very close to, the
surface of a sphere, and yield virtually the same polyhedral solutions.
We also discussed a field theory in which polyhedra arise in
a slightly different context, as structures around which the 
charge density of a soliton solution is localized, and mentioned an example
of how a comparison with point particle systems can be used to gain 
some insight.

Even very simple interactions can yield complex structures 
and surprising results. Although the use of modern computers  
is vital in determining minimal energy arrangements, often a 
geometrical point of view can lead to new insights. As we have seen, on
many occasions the geometry can be universal and progress made in one
area can be directly applied to several different situations, leading
to advances in a wide range of apparently unrelated disciplines.

\section*{Acknowledgements}
PMS acknowledges the EPSRC for an advanced fellowship.\\

\end{document}